\documentclass[12pt]{article}
\usepackage{times}
\usepackage{geometry}
\usepackage[utf8]{inputenc}
\usepackage{enumitem,amssymb}
\usepackage{ragged2e}
\usepackage{pifont}
\usepackage[sort,super]{natbib}
\usepackage{epsfig}
\usepackage{wrapfig}
\usepackage{url}
\usepackage[colorlinks=true, allcolors=blue]{hyperref}
\usepackage{color}
\usepackage{amsmath,amsfonts}
\usepackage{graphicx}
\usepackage[table]{xcolor}
\usepackage{multirow}
\usepackage{multicol}
\usepackage{tabu}
\usepackage{array}
\newcolumntype{C}[1]{>{\centering\let\newline\\\arraybackslash\hspace{0pt}}m{#1}}

\begin{document}
\clearpage
\thispagestyle{empty}
\raggedright

\Large Astro2020 Science White Paper \linebreak

\huge Energetic Particles of Cosmic Accelerators I: \linebreak
\ \linebreak
\textit{\Large Galactic Accelerators}\linebreak
\normalsize

\noindent \textbf{\underline{Thematic Areas}} \linebreak
\noindent \textit{Primary: Multi-Messenger Astronomy and Astrophysics}\linebreak
\noindent \textit{Secondary: Stars and Stellar Evolution}\linebreak



\textbf{Principal Authors:}

Name: Tonia M. Venters\\ 
Institution: NASA Goddard Space Flight Center\\ 
Email: \href{mailto:tonia.m.venters@nasa.gov}{tonia.m.venters@nasa.gov}\\ 
Phone: 301-614-5546\\
\bigskip
Name: Kenji Hamaguchi\\ 
Institution: NASA GSFC/CRESST/University of Maryland, Baltimore County\\
Email: \href{mailto:kenji.hamaguchi@nasa.gov}{kenji.hamaguchi@nasa.gov}\\
\bigskip

\textbf{Co-authors:}\\
Terri J. Brandt (NASA GSFC), Marco Ajello (Clemson University), Harsha Blumer (West Virginia University), Michael Briggs (University of Alabama, Huntsville), Paolo Coppi (Yale University), Filippo D'Ammando (INAF Istituto di Radioastronomia), Micha\"{e}l De Becker (University of Li\`{e}ge, Belgium), Brian Fields (University of Illinois, Urbana-Champaign), Sylvain~Guiriec (George Washington University/NASA GSFC), John W. Hewitt (University of North Florida), Brian Humensky (Columbia University), Stanley~D.~Hunter (NASA GSFC), Hui~Li (LANL), Amy Y. Lien (NASA GSFC/CRESST/UMBC), Francesco Longo (University of Trieste/INFN Trieste), Alexandre~Marcowith (Laboratoire Univers et Particules de Montpellier), Julie McEnery (NASA GSFC), Roopesh~Ojha (UMBC/NASA GSFC), Vasiliki Pavlidou (University of Crete), Chanda Prescod-Weinstein (University of New Hampshire), Marcos~Santander (University of Alabama), John A. Tomsick (UC Berkeley/SSL), Zorawar~Wadiasingh (NASA/GSFC), Roland~Walter (University of Geneva)\\
  
\pagebreak
\justifying
\setcounter{page}{1}

\section*{Executive Summary} 
The high-energy universe has revealed that energetic particles are ubiquitous in the cosmos and play a vital role in the cultivation of cosmic environments on all scales. Our pursuit of more than a century to uncover the origins and fate of these cosmic energetic particles has given rise to some of the most interesting and challenging questions in astrophysics. 
%
%
Energetic particles in our own galaxy, galactic cosmic rays (GCRs), engage in a complex interplay with the interstellar medium and magnetic fields in the galaxy, giving rise to many of its key characteristics. For instance, GCRs act in concert with galactic magnetic fields to support its disk against its own weight. GCR ionization and heating are essential ingredients in promoting and regulating the formation of stars and protostellar disks. 
GCR ionization also drives astrochemistry, leading to the build up of complex molecules in the interstellar medium. 
GCR transport throughout the galaxy generates and maintains turbulence in the interstellar medium, alters its multi-phase structure, and amplifies magnetic fields.
GCRs could even launch galactic winds that enrich the circumgalactic medium and alter the structure and evolution of galactic disks.

As crucial as they are for many of the varied phenomena in our galaxy, there is still much we do not understand about GCRs. While they have been linked to supernova remnants (SNRs), it remains unclear whether these objects can fully account for their entire population, particularly at the lower ($\lesssim 1$ GeV per nucleon) and higher ($\sim$ PeV) ends of the spectrum. In fact, it is entirely possible that the SNRs that have been found to accelerate CRs merely \textit{re-accelerate} them, leaving the origins of the original GCRs a mystery.
The conditions for particle acceleration that make SNRs compelling source candidates are also likely to be present in sources such as protostellar jets, superbubbles, and colliding wind binaries (CWBs), but we have yet to ascertain their roles in producing GCRs. 
For that matter, key details of diffusive shock acceleration (DSA) have yet to be revealed, and it remains to be seen whether DSA can adequately explain particle acceleration in the cosmos. 

This White Paper is the first of a two-part series highlighting the most well-known high-energy cosmic accelerators and contributions that MeV $\gamma$-ray astronomy will bring to understanding their energetic particle phenomena.
%
%
For the case of GCRs, MeV astronomy will:
\begin{itemize}
    \item[1.] Search for fresh acceleration of GCRs in SNRs; 
    \item[2.] Test the DSA process, particularly in SNRs and CWBs;
    \item[3.] Search for signs of CR acceleration in protostellar jets and superbubbles.
\end{itemize}

\vspace{-2.0ex}
\section*{Supernova Remnants}
\vspace{-2.0ex}

Supernova remnants (SNRs) have long been suspected of being the sources of GCRs. Initially, this proposal was motivated by SN energetics -- assuming the typical kinetic energy of $\sim 10^{51}$ ergs and an event rate of $\sim 1$ per century, SNe could supply the observed energy density of GCRs if $\sim$10\% of the explosion energy could be harnessed. SN blast waves provide a natural site for particle acceleration via DSA -- CRs diffuse back and forth across the SN shock, gaining energy with each crossing, resulting in an $\sim E_{\rm CR}^{-2}$ that with subsequent propagation and energy losses becomes the $\sim E_{\rm CR}^{-2.7}$ spectrum that we observe on Earth \cite{1978MNRAS.182..147B}. The success of this paradigm in reproducing the measured CR spectrum, energy density, and composition is the reason for the wide acceptance of SNRs 
\begin{wrapfigure}{r}{0.625\textwidth}
\begin{center}
\vspace{-2.5ex}
\resizebox{4.in}{!}
{
\includegraphics[trim = 77mm 75mm 75mm 45mm, clip]{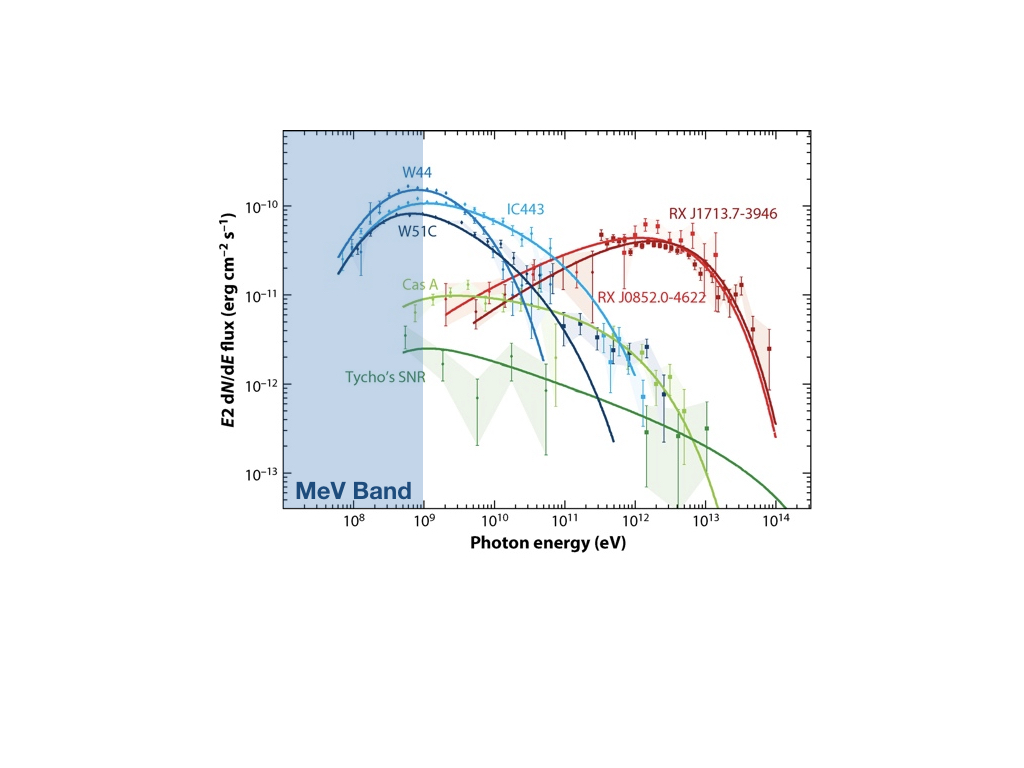}}
\vspace{-1.5ex}
\caption{\small \textit{$\gamma$-ray spectra of several SNRs \cite{Funk:2015ena}. For young SNRs such as Tycho's SNR, we need a sensitive MeV telescope to identify a pion bump that would indicate CR acceleration.} \label{fig:ic443_image}}
\vspace{-5.0ex}
\end{center}
\end{wrapfigure}
as the likely sources of GCRs even though compelling evidence has remained elusive for many decades. Verification that SNRs accelerate CRs was finally realized 
when \textit{Fermi}-LAT and AGILE detected $\gamma$-rays linked to the decay of $\pi^{0}$'s (the so-called ``pion bump'') produced through inelastic collisions between CRs and the ambient medium \cite{2011ApJ...742L..30G,ack13snrpionbump}.

While compelling, $\gamma$-ray observations of SNRs in the GeV and TeV bands have reinvigorated the debate over their capabilities in accelerating CRs. For starters, the CR spectra required to fit GeV and TeV $\gamma$-ray spectra are steeper at high energies than the CR spectrum predicted by standard non-relativistic DSA \cite{1978MNRAS.182..147B,1978ApJ...221L..29B}. Additionally, many of the SNRs observed in $\gamma$-rays, including W44 and IC 443, are middle-aged ($t_{\rm age} \sim 10^4$ yrs) with slow shocks ($v_{\rm sh} \lesssim 100$ km$/$s) for which particle acceleration is expected to be relatively inefficient, particularly at the high-energy end of the GCR spectrum, which is presumably $\sim$ PeV\cite{2012arXiv1206.5018D,Lee:2015ima}.
Accommodating these observations requires either re-acceleration of pre-existing GCRs \cite{2010ApJ...723L.122U,Lee:2015ima} and/or modifications to the DSA paradigm. 
The faster shocks in younger SNRs, such as Tycho's SNR ($t_{\rm age} \sim 450$ yrs), could efficiently accelerate fresh particles to CR energies, but as yet, the ``pion bump'' has not been found in these SNRs (see Fig.~\ref{fig:ic443_image}). Furthermore, as the elemental abundances of GCRs are different from those of the interstellar medium \cite{Murphy:2016kyv}, CR re-acceleration in older SNRs would require revisions in our understanding of CR transport throughout the galaxy in order to reproduce measured CR elemental abundances \cite{2018JHEAp..19....1D}. If young SNRs do accelerate CRs, the seed particles likely include both stellar wind material accelerated in the forward shock and supernova ejecta accelerated in the reverse shock \cite{2013A&A...552A.102T}. In that case, we can expect differences in their elemental abundances with respect to those of older SNRs, but \textit{without MeV astronomy, we have no way of measuring these elemental abundances in situ.}

Beyond the possibility of GCR re-acceleration, the $\gamma$-ray observations of SNRs could signify the need for modifications to the DSA paradigm. In efficient acceleration, a significant fraction of the SN energy is supplied to the accelerated CRs; hence, far from behaving as inactive test particles, these CRs will exert a pressure force, feeding back on the incoming plasma flow and modifying SN shock structure and dynamics \cite{2012APh....35..300T,2012JCAP...07..038C}. Including the backreaction of CRs in nonlinear DSA theories have resulted in concave spectra that are actually \textit{flatter} at high energies, rather than the steeper spectra exhibited in $\gamma$-rays in some cases\cite{2012JCAP...07..038C}. However, in addition to exerting pressure, CR streaming along magnetic field lines can excite plasma waves that will amplify the magnetic field, but will also drain energy from the CRs leading to steep spectra \cite{2012JCAP...07..038C}. The nature of the medium surrounding the SNR likely plays a role in the DSA process as the density and ionization of the medium substantially impact how and when accelerated particles escape upstream of the SN shock resulting in steep spectra \cite{2011MNRAS.415.1807D,2011NatCo...2E.194M,2012ApJ...755..121B}, high-energy $\gamma$-ray and neutrino emission in nearby molecular clouds being bombarded by CRs escaping the SNR \cite{2010PASJ...62..769C,2011MNRAS.415.1807D,2013A&A...552A.102T}, and time-varying release of CRs of a given energy over the course of the SNR's evolution \cite{2011MNRAS.415.1807D,2012APh....35..300T,2013A&A...552A.102T}. SNR reverse shocks can also accelerate particles leading to time evolution in the spectra and morphologies of their nonthermal emission \cite{2013A&A...552A.102T}. Combining MeV observations with those at GeV and TeV energies from current and next-generation observatories such as \textit{CTA} will allow us to more significantly constrain DSA theories.


Further insight into SNR CR populations requires elemental abundance measurements \textit{at the source} (\textit{i.e.}, without propagation effects), a unique purview of MeV $\gamma$-ray astronomy as elements can be identified through nuclear de-excitation lines that appear in the MeV band. As such, MeV spectral line measurements will be crucial in assessing the role of re-acceleration in older SNRs and in providing much needed information about the environments of younger SNRs. Continuum measurements at MeV energies in young SNRs will also be needed to identify the ``pion bump'' and verify that they do indeed accelerate fresh CRs. MeV measurements of bremsstrahlung emission in SNRs will determine their electron energy densities, that when combined with radio measurements from instruments such as \textit{LOFAR} or \textit{SKA} of synchrotron emission, will determine their magnetic field strengths, a crucial input for the DSA process \cite{2018JHEAp..19....1D}. Additional tests of the DSA process will be enabled by MeV astronomy as observations of  SNRs will allow for population studies that will determine the evolution of the high-energy spectrum and morphology with SNR age. This will also provide a means for assessing the impact(s) of the circumremnant medium and the reverse shock, particularly when combined with multiwavelength observations in radio, X-rays, and GeV and TeV $\gamma$-rays. MeV astronomy will also provide an excellent synergy with next-generation neutrino observatories (\textit{e.g.}, \textit{IceCube-Gen 2}) as hadronic interactions occurring within in SNRs and nearby molecular clouds produce neutrinos as well as $\gamma$-rays.




\vspace{-3.0ex}
\section*{Protostellar Jets and Superbubbles}
\vspace{-2.0ex}

MeV astronomy will also pave the way for searching for CR acceleration in the earlier life stages of massive stars, as well as in death. 
Shocks formed at the surface of an accreting massive protostar or at the interface between protostellar jets and the ambient medium may also provide the conditions necessary for accelerating CRs \citep{gaches18,romero10} that in turn drive complex chemistry by ionizing dense molecular gas where ultraviolet radiation cannot penetrate \citep{grenier15,gaches18}. Interactions of accelerated CRs within the molecular cloud will produce copious amounts of MeV $\gamma$-rays that will provide insight into star formation processes and the nature of the protostar environment \citep{romero10}. Other exquisite test beds for CR physics (motivated by GCR composition measurements\cite{Murphy:2016kyv}) are available in the form of superbubbles\cite{2014A&ARv..22...77B}, such as the Cygnus X superbubble, that have been blown out by multiple powerful winds from young massive stars and/or OB associations and multiple supernovae. The ensemble of shocks resulting from winds and supernovae can be relatively efficient in accelerating or re-accelerating CRs over extended periods of time ($\sim 10$ Myr). The turbulent medium resulting from vigorous star formation activity in superbubbles also provides a unique opportunity to study CR propagation through highly turbulent media and to study their impact on the superbubble environment. MeV spectral measurements will be needed to extend the $\gamma$-ray spectrum of Cygnus cocoon measured by the \textit{Fermi}-LAT \citep{Ackermann:2011lfa} to lower energies, allowing for distinction between emission from CR nuclei and CR electrons and providing measurements of their respective energy densities \citep{2018JHEAp..19....1D}. Similarly, it has been suggested that the \textit{Fermi} Bubbles may be the result of CRs accelerated in the combined wind from SN explosions of massive stars (the Starburst Scenario) \citep{2010ApJ...711..818S}. MeV spectral measurements of the \textit{Fermi} Bubbles in the MeV band, particularly at energies below the ``pion bump'', will determine whether the $\gamma$-ray continuum can be explained by Inverse Compton emission from high-energy electrons (as in an AGN-like scenario) or if hadronic emission such as that expected in a Starburst scenario is needed.

\vspace{-3.0ex}

\section*{Colliding Wind Binaries}
\vspace{-2.0ex}

\begin{wrapfigure}{r}{0.525\textwidth}
\begin{center}
\vspace{-4.5ex}
\includegraphics[height = 2.5in]{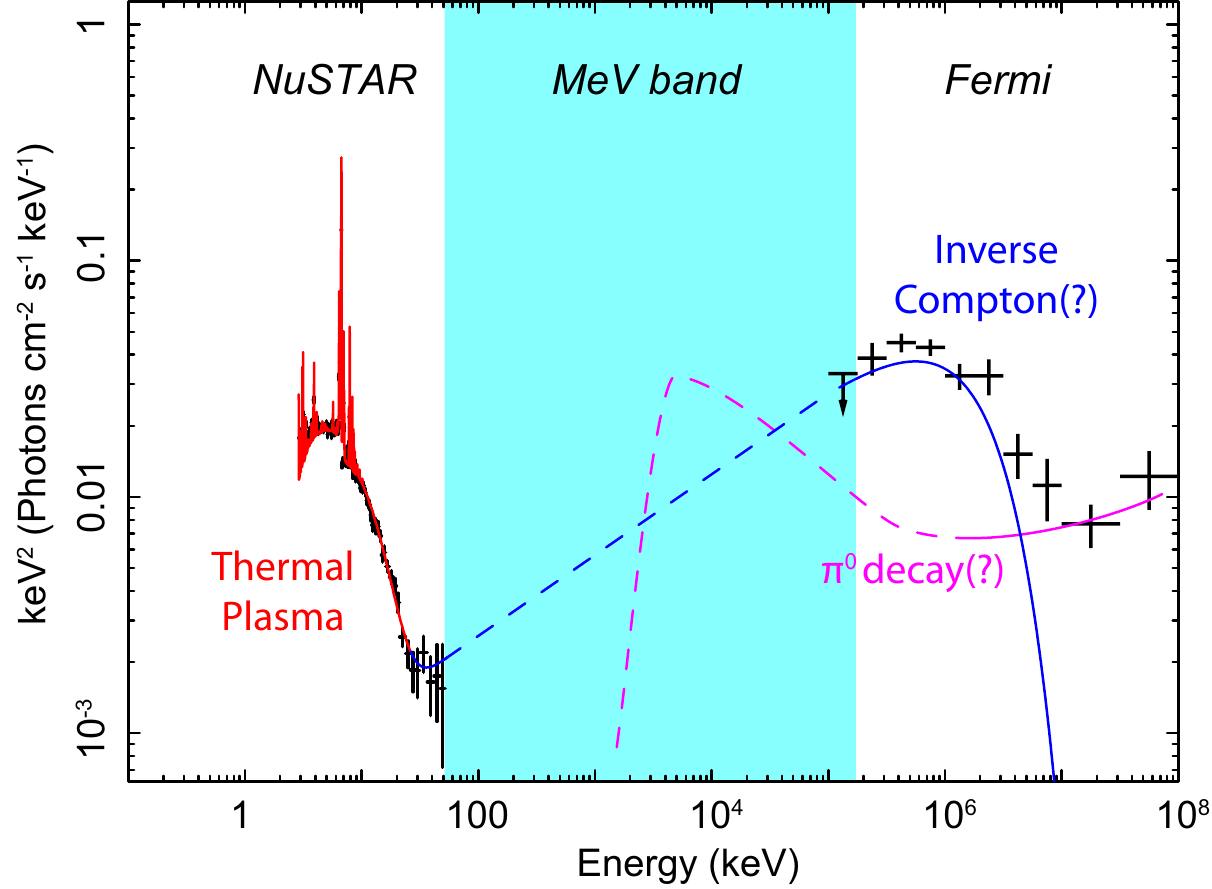}
\vspace{-1.5ex}
    \caption{\small NuSTAR \textit{and} Fermi\textit{-LAT spectra of $\eta$Car \citep{hamaguchi18}. We need a sensitive MeV telescope to identify the signs of CR acceleration (notional $\pi^{0}$ spectrum provided for comparison.)}
    \label{fig:etacar_spectrum}}
    \vspace{-5.0ex}
\end{center}
\end{wrapfigure}
More than half of massive stars form a 
binary system with another massive star. UV 
radiation in the two stars drives 
strong winds that collide, resulting in persistent shocks that accelerate particles to high energies via DSA.
The accelerated particles would not be so energetic as those produced in supernova remnants or active galactic nuclei,
but the binary systems last longer and thus may contribute more to the low-energy side of the cosmic-ray spectrum\cite{2017A&A...600A..47D}.
The shock condition --- the velocity and density of pre-shock gases --- can be estimated from stellar wind parameters and binary orbital solution. CWBs are good laboratories for the shock acceleration physics.

The wind-wind collision produces hot plasma of $T \sim 20$$-$$40$~MK, whose thermal emission dominates the soft X-ray band. As such, searches for non-thermal emission have mostly been conducted at energies above $\sim 10$ keV, but it had remained elusive for decades due to lack of sensitivity in this band. The unprecedented high-energy sensitivity of \textit{NuSTAR} and \textit{Fermi} enabled confirmation of non-thermal emission for the first time in $\eta$Carinae,
%
%
an enigmatic supermassive CWB system\cite{Abdo2010,hamaguchi18} (see Fig. \ref{fig:etacar_spectrum}).
The emission detected with both \textit{NuSTAR} and \textit{Fermi} likely originates from Inverse Compton scattering of 
stellar UV photons 
by relativistic electrons accelerated at the colliding wind shocks.
It disappeared around periastron probably caused by variation of particle acceleration efficiency in its long eccentric, binary orbit.
The Fermi observatory might also found evidence for hadron acceleration in eta Carinae, pion-decay through interactions 
between accelerated hadrons and circumstellar materials.
These observatories also found hints of non-thermal emission from more massive CWB systems (e.g., WR140, $\gamma^{2}$~Vel). Thus, particle acceleration activity may actually occur in the CWB systems.

Although the \textit{NuSTAR} and \textit{Fermi} observations confirmed the presence of relativistic electrons in $\eta$Carinae, the shock acceleration mechanism cannot be understood without the MeV $\gamma$-ray information. The $\eta$Carinae spectrum in Fig.~\ref{fig:etacar_spectrum} has a huge gap between 0.05-100 MeV, where the Inverse Compton emission in the hard X-ray and GeV $\gamma$-ray bands likely connects and the $\pi^{0}$-decay component is possibly more prominently seen. Sensitive MeV $\gamma$-ray data are crucially needed to unambiguously separate these two components at each binary orbit and to understand how hadronic and leptonic particles are accelerated in the wind collision region. In addition, even after 10 years of observing the entire sky with \textit{Fermi}, $\eta$Carinae is the only $\gamma$-ray confirmed particle acceleration site among the known CWB systems (synchrotron radio emission has indicated particle acceleration in other CWBs\cite{2013A&A...558A..28D}). This indicates that most CWBs are faint in GeV $\gamma$-rays, and may also mean that they do not accelerate particles to GeV energies. A sensitive all-sky survey in the MeV band is crucial to understand particle acceleration in CWB systems.

\vspace{-3.5ex}

\section*{The Promise of the Next Decade}

\vspace{-2.0ex}

The 2020s will usher in an new era of MeV astronomy that will revolutionize our quest to understand the most energetic particles and phenomena in our universe (for more details and many more examples, see \textit{e.g.}, the \textit{e-ASTROGAM} White Book \cite{2018JHEAp..19....1D}). In order to make the measurements necessary for the particle accelerators in our galaxy, we need an MeV telescope with excellent continuum sensitivity that bridges the gap between the thermal and non-thermal regimes. We also need a telescope with excellent energy resolution in order to determine elemental abundances via nuclear line measurements. Possible MeV missions include: \textit{AdEPT}\cite{2014APh....59...18H}, \textit{AMEGO}\cite{amego_2018_aa}, \textit{e-ASTROGAM}\cite{2018JHEAp..19....1D}, \textit{COSI}\cite{Kierans:2017bmv}, and \textit{SMILE}\cite{2015ApJ...810...28T}. This science will also leverage synergies with instruments currently operating in the GeV band, such as \textit{Fermi} and \textit{AGILE}, next-generation hard X-ray telescopes, such as \textit{HEX-P} or \textit{FORCE}, next-generation TeV telescopes, such as \textit{CTA}, and next-generation neutrino observatories, such as \textit{IceCube-Gen 2}.


\begin{table}[h]
    \centering
    \begin{tabular}{|C{2.0in}|C{2.4in}|C{1.6in}|}
    \hline
    \rowcolor{lightgray} \multicolumn{3}{|c|}{GCR Science with MeV $\gamma$-rays}\\
    \hline
    Science Objective & Strategy & Instruments \\
    \hline
    \centering \multirow{3}{2.0in}{\raggedright Distinguish between fresh CR acceleration and CR re-acceleration in SNRs} & \raggedright Search for low-energy cutoff consistent with $\pi^0$-decay in the MeV spectra of SNRs & \multirow{4}{1.4in}{\small \centering \textit{AdEPT}, \textit{AMEGO}, \textit{e-ASTROGAM}, \textit{COSI}, \textit{SMILE}} \\ 
    \cline{2-2}
    & \raggedright Search for MeV lines that would allow elemental abundance measurements & \\ 
    \cline{1-3}
    \multirow{1}{2.0in}{\raggedright Distinguish between leptonic and hadronic processes} & \raggedright Search for low-energy cutoff consistent with $\pi^0$-decay & \multicolumn{1}{c|}{\begin{minipage}{1.4in}\centering{\small \textit{AdEPT}, \textit{AMEGO}, \textit{e-ASTROGAM},  \textit{COSI}, \textit{SMILE}}\end{minipage}} \\
    \cline{2-3}
    & \raggedright Search for neutrinos & IceCube-Gen 2 \\
    \cline{1-3}
    \raggedright Characterize CR transport throughout the galaxy & \raggedright Measure elemental abundances in energetic galactic sources & \multicolumn{1}{c|}{\begin{minipage}{1.4in}\centering{\small \textit{AdEPT}, \textit{AMEGO}, \textit{e-ASTROGAM}}, \textit{COSI}, \textit{SMILE}\end{minipage}}\\ 
    \cline{1-3}
    \multirow{7}{2.0in}{\raggedright Test the DSA process} & \raggedright Measure electron energy densities that would allow measurements of magnetic field strengths & \multirow{7}{1.4in}{\small \centering \textit{AdEPT}, \textit{AMEGO}, \textit{e-ASTROGAM}, \textit{COSI}, \textit{SMILE}, \textit{Fermi}-LAT, \textit{AGILE}, \textit{CTA}, \textit{FORCE}, \textit{HEX-P}, LOFAR, SKA}\\ 
    \cline{2-2}
    & \raggedright Perform population studies of SNR in order to study time evolution of nonthermal emission spectra and morphology & \\
    \cline{2-2}
    & \raggedright Search for $\gamma$-ray emission from nearby molecular clouds & \\
    \hline
    \centering \multirow{3}{2.0in}{\raggedright Search for CR acceleration in jets and superbubbles} & \raggedright Search for and measure MeV spectra & \multirow{3}{1.4in}{\small \centering \textit{AdEPT}, \textit{AMEGO}, \textit{e-ASTROGAM},  \textit{COSI}, \textit{SMILE}} \\
    \cline{2-2}
    & \raggedright Search for MeV emission from surrounding molecular clouds & \\
    \hline
    \raggedright Decipher the origins of \textit{Fermi} Bubbles & \raggedright Measure IC spectrum in the MeV band  & \multicolumn{1}{c|}{\begin{minipage}{1.in}\centering{\small \textit{AdEPT}, \textit{AMEGO}, \textit{e-ASTROGAM},  \textit{COSI}, \textit{SMILE}}\end{minipage}} \\
    \hline
    \end{tabular}
\end{table}

\pagebreak
\bibliography{cosacc}

\end{document}